\documentstyle[aps,epsf,epsfig]{revtex}

\begin{document}
\author{D. Foerster}
\address{CPMOH, UMR 5798, Universit\'{e} de Bordeaux I \\
351, cours de la Lib\'{e}ration, F - 33405 Talence Cedex}
\author{A.A. Ovchinnikov}
\address{Max-Planck-Institut f\"{u}r Physik komplexer Systeme \\
N\"{o}thnitzer Str. 38, D-01187 Dresden, Germany \\
and \\
Joint Institute for Chemical Physics of the Russian \\
Academy of Sciences,\\
117977, Moscow, Russia}
\title{The emergence of charged collective modes from a \\
large $N$ extrapolation of the Hubbard model}
\maketitle

\begin{abstract}
We consider a symplectic extrapolation of the Hubbard model of $N$ fold
replicated electrons and solve this model exactly in two special cases, at $%
N=\infty $ in the bosonic sector and for any $N$ on a dimer of two points.
At $N=\infty $ we find a multiplet of collective modes that contains neutral
spin fluctuations and charged pair fluctuations that are degenerate with
each other at zero doping. Our solution of the symplectic model on a dimer
of two points for any $N$ interpolates smoothly between $N=1$ and $N=\infty $
without any visible discontinuity. These results suggest that the inclusion
of charged pairing modes in weakly doped antiferromagnets is essential and
that an expansion about the $N=\infty $ limit is appropriate in this context.
\end{abstract}

\section{Introduction and Motivation}

Calculational schemes that are perturbative in the Coulomb interaction are
known to be inadequate for describing strongly interacting electrons in
systems such as, for example, transition metal oxides, organic conductors
and quantum dots \cite{Tokura} and this necessitates the use of non
perturbative methods, such as the renormalization group, dynamical mean
field theory \cite{DMFT} and various numerical methods. The $1/N$ expansion
for $N$ fold replicated degrees of freedom (with subsequent extrapolation to
the physical value of $N=2$ or $N=3$) was successfully applied to magnetic
impurities \cite{impurities} and in other areas of physics \cite{Brezin}. In
almost all applications to electronic systems, this expansion uses
semiclassical saddle point integration and its success in the case of the
repulsive Hubbard model \cite{Affleck} is open to question because it is
unclear where a suitable saddle point exists. An alternative version of the $%
1/N$ expansion was discovered by 't Hooft \cite{tHooft} in the context of $%
SU(N)$ Quantum Chromodynamics and relies instead on dominance of diagrams of
simple topology for $N\rightarrow \infty $. Recently it was noticed \cite{DF}
that 't Hooft's topological classification of diagrams can be applied to the
Hubbard model, with collective excitations playing the role that gauge
bosons play in Quantum Chromodynamics. This raises the challenge of
reconstructing the large $N$ extrapolation for electron systems from the
very beginning in a way that avoids saddle point integration.

The symplectic extrapolation of the Hubbard model \cite{Sachdev},\cite{DF}
is interesting because, unlike the $SU(N)$ extension, it conserves the
notion of two component spins and permits pairing. In the present paper, we
solve this model exactly in two special cases (i) in the bosonic sector on
an infinite lattice at $N=\infty $ and (ii) on a dimer of two points for any 
$N$. Our solution at $N=\infty $ shows the existence not only of neutral
spin fluctuations but also of charged pair fluctuations that are in the same
multiplet and exactly degenerate with the spin fluctuations at zero doping.
Our solution of the dimer for any $N$ indicates a smooth interpolation
between $N=1$ and $N=\infty $ without any apparent singularity. Taken
together these results indicate that (i) both spin and pair fluctuations
must be included from the very beginning in any theory of weakly doped
antiferromagnets and (ii) that the expansion in powers of $1/N$ about an
appropriate $N=\infty $ limit is a legitimate procedure in this context.

The symplectic extrapolation of Hubbard's model \cite{Hubbard} that we
consider here is based on $U(N,q)$, the group of unitary transformations on $%
N$ quaternions and it leads to the following Hamiltonian \cite{DF}: 
\begin{eqnarray}
H_{U(N,q)} &=&\frac{U}{N}\sum_{x}\left[ \left\{ \left( \psi _{x}\varepsilon
\psi _{x}\right) ^{+}\left( \psi _{x}\varepsilon \psi _{x}\right) \right\}
_{symm}+\left( \psi _{x}^{+}\psi _{x}\right) ^{2}\right] +\sum_{x,y}t_{xy}%
\psi _{x}^{+}\psi _{y}  \label{symplectic} \\
\psi _{x}\varepsilon \psi _{x} &=&\sum_{i=1..N,\alpha \beta =1,2}\psi
_{x\alpha i}\left( i\sigma _{y}\right) _{\alpha \beta }\psi _{x\beta i}%
\mbox{,\ \ \ }\psi _{x}^{+}\psi _{x}=\sum_{i=1..N,\alpha =1,2}\psi _{x\alpha
i}^{+}\psi _{x\alpha i}  \nonumber
\end{eqnarray}
Here $\left\{ {}\right\} _{symm}$ denotes symmetrization and the electrons $%
\psi _{x\alpha i}$ occur in $N$ copies with $i=1..N$ but we will often
suppress this extra index. For our purposes invariance of $H_{U(N,q)}$ under 
$SU(N)$ transformations is sufficient and $\psi _{x}\varepsilon \psi _{x}$
and $\psi _{x}^{+}\psi _{x}$ in eq(\ref{symplectic}) are indeed invariant
under $SU(N)$ because $\psi _{x1i}$ and $\psi _{x2i}$ transform, by fiat,
according to mutually conjugate $SU(N)$ transformations. If we imagine, for
the moment, that the interactions in eq(\ref{symplectic}) come about via
exchange of suitable bosons, then both neutral and charged bosons must be
exchanged. It will be shown below that this naive expectation is correct and
that the symplectic extrapolation of the Hubbard model contains indeed both
neutral and charged collective excitations.

This paper is organized as follows. In section 2 we solve the $SU(N)$
Hubbard model first for two points and any $N$ using $SU(2)$ pseudo spins
and we also find the $SU(\infty )$ Hamiltonian in the bosonic sector for an
infinite lattice. In section 3 we determine the $N\rightarrow \infty $
asymptotics of the symplectic model after mapping it onto a model of the
double exchange type and we also solve the symplectic model on a dimer for
any $N$ using the group $SP(4)\sim SO(5)$ as a classifying symmetry. In
section 4 we identify the spectrum of the asymptotic $SU(\infty )$ and $%
U(\infty ,q)$ Hamiltonians. Section 5 serves to counterbalance our extensive
use of ancient boson operator methods by reformulating our problem in
standard path integral language. Our conclusions are given in section 6.

\section{Exact results for limiting cases of the SU(N) Hubbard model}

\subsection{The SU(N) Hubbard model for a dimer}

For simplicity, we begin with an $SU(N)$ extrapolation \cite{Affleck} of the
Hubbard model \cite{Hubbard}: 
\begin{equation}
H_{SU(N)}=\frac{U}{2N}\sum_{x}n_{x}{}^{2}+\sum_{x,y}t_{xy}\psi _{x}^{+}\psi
_{y}  \label{SUN}
\end{equation}
which we simplify even further by specializing it to a dimer of only two
points :

\begin{equation}
H_{SU(N)}^{\mbox{dimer}}=\frac{U}{2N}\left( n_{1}{}^{2}+n_{2}{}^{2}\right) +%
\frac{1}{2}\sum_{i=1..N}\psi _{1i}^{+}\psi _{2i}+\psi _{2i}^{+}\psi _{1i}
\label{SunDimer}
\end{equation}
Here $n_{a}=\sum_{i=1..N}\psi _{ai}^{+}\psi _{ai}$, $a=1,2$ and $i=1..N$
enumerates the electron species and we have chosen hopping amplitudes with a
(naive) bandwidth of 1. This Hamiltonian describes electrons on a ladder of
length $N$ with hopping along the rungs $i$ or tunnelling between two
mesoscopic quantum dots on a semiconducting substrate, each doped with $N$
electron donors. The position $\alpha =1,2$ on the dimer can be viewed as a
pseudo spinor index and $H_{SU(N)}^{\mbox{dimer}}$ can be expressed entirely
in terms of pseudo spin operators as follows: 
\begin{eqnarray}
H_{SU(N)}^{\mbox{dimer}} &=&\frac{U}{N}T_{z}^{2}+T_{x}+const  \nonumber \\
\overrightarrow{T} &=&\frac{1}{2}\sum_{i=1..N,\alpha =1,2}\psi _{ai}^{+}%
\overrightarrow{\tau }_{\alpha \beta }\psi _{\beta i}
\end{eqnarray}
where the constant has physical meaning because it depends on electron
filling. Because the Hamiltonian decomposes into blocks of distinct pseudo
spins we may diagonalize it separately in each block and determine the
optimal value of the pseudo spin at the end (it is impossible to diagonalize
the $SU(N)$ dimer for large $N$ by brute force because the dimension of its
Hilbert space explodes like $4^{N}$). It it is useful to diagonalize the
hopping term by interchanging the conventional representations of $T_{z}$
and $T_{x}$: 
\begin{equation}
H_{SU(N)}^{\mbox{dimer}}\rightarrow \frac{U}{N}T_{x}^{2}+T_{z}+const
\label{interchanged}
\end{equation}
Because $\frac{U}{N}\rightarrow 0$ as $N\rightarrow \infty $ a naive first
approximation to the ground state is $|s,s_{z}=-s>$. To understand the
action of the Hamiltonian on states in the vicinity of $|s,-s>$, we express
it in terms of Holstein-Primakoff \cite{Holstein-Primakoff} oscillators:

\begin{eqnarray}
T_{+} &=&a^{+}\sqrt{2s-a^{+}a}\mbox{, \ \ \ \ }T_{-}=\sqrt{2s-a^{+}a}a%
\mbox{, \ \ \ \ }T_{z}=-2s+a^{+}a  \label{Holstein} \\
H_{SU(N)}^{\mbox{dimer}} &=&\frac{1}{2}\left( 
\begin{array}{cc}
a^{+} & a
\end{array}
\right) \left( 
\begin{array}{cc}
\frac{Us}{N}+1 & \frac{Us}{N} \\ 
\frac{Us}{N} & \frac{Us}{N}+1
\end{array}
\right) \left( 
\begin{array}{c}
a \\ 
a^{+}
\end{array}
\right) +const+O\left( \frac{1}{N}\right)  \nonumber
\end{eqnarray}
and then diagonalize it with the help of a Bogoljubov transformation: 
\begin{eqnarray}
H &=&\sqrt{1+\frac{2Us}{N}}b^{+}b+const+O\left( \frac{1}{N}\right)
\label{Bogoljubov} \\
b &=&a\cosh \theta +a^{+}\sinh \theta \mbox{, \ \ \ }\tanh 2\theta =\frac{%
\frac{Us}{N}}{1+\frac{Us}{N}}  \nonumber \\
|G &>&=\exp (-\frac{\tanh \theta }{2}a^{+}a^{+})|0>  \nonumber
\end{eqnarray}
The use of the Holstein-Primakoff expansion for $N\rightarrow \infty $ is
justified, a posteriori, because the number of $a^{+}a^{+}$ pairs converges
in the ground state. The optimal value of the total spin $s$ is the maximal
one that is permitted at a given filling e.g. $s=\frac{N}{2}$ at half
filling \cite{SUNinvariance}.

By numerically diagonalizing the $SU(N)$ dimer for finite $N$ the
convergence to its $N=\infty $ limit can be checked. We consider the gap $%
\Delta _{N}(U)$ between the ground state and the first excited state and
which, according to eq(\ref{Bogoljubov}) should converge (at half filling)
to $\sqrt{1+U}$ and our calculation confirms this convergence. Figure {\bf 1 
} displays the gap $\Delta _{N}(U)$ for $N=2...5$ and shows that the
convergence with $N$ is non uniform in $U$. This reflects the failure of the
limits $N\rightarrow \infty $ and $U\rightarrow \infty $ to commute that
follows from eq(\ref{SUN}). We therefore expect the extrapolation to the
physical value $N=2$ to run into difficulties if the Coulomb interaction is
too large compared with the bandwidth.

\begin{center}
\begin{figure}[h]
\mbox{\epsfig{file=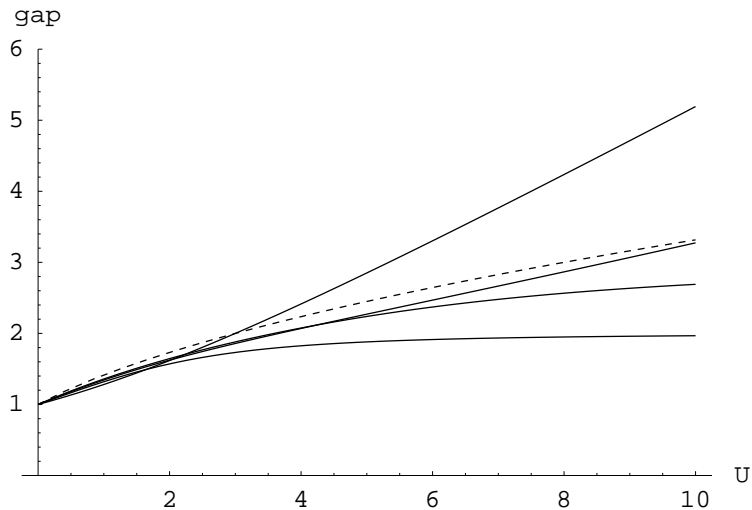,width=10cm}}\\
\caption{ Oscillating convergence in $N$ of the gaps $\Delta _{N}(U)$ of the
SU(N) dimer for $N=2..5$ towards $\Delta _{\infty}(U)$. The highest curve is
for $N=2$, the lowest for $N=3$ and the asymptotic one is dotted.}
\end{figure}
\end{center}

\subsection{Hamiltonian of the SU($\infty $) Hubbard model on any lattice}

We now generalize the operator representation of the $SU(N)$ dimer to an
arbitrary lattice. Because we want to use the notion of a Fermi surface, we
transform the $SU(N)$ Hamiltonian of eq(\ref{SUN}) to plane waves:

\begin{equation}
H_{SU(N)}=\frac{U}{2Nvol}\sum_{p-q=r-s}B_{pq}^{+}B_{rs}+\sum e_{p}B_{pp}
\label{Fourier}
\end{equation}
with $\psi _{x}=\frac{1}{\sqrt{vol}}\sum e^{ipx}\psi _{p}$, $\ B_{pq}\equiv
\psi _{p}^{+}\psi _{q}$ \ \ and $e_{p}\equiv \sum t_{x0}e^{ipx}$. To find
the $1/N$ expansion of the $B_{pq}$ operators we examine the commutators of
their fluctuating components: 
\begin{eqnarray}
\left[ B_{pq},B_{rs}\right] &=&\delta _{qr}B_{ps}-\delta _{ps}B_{qr}
\label{heuristics} \\
&=&\delta _{qr}\left( B_{ps}-N\delta _{ps}n_{p}\right) -\delta _{ps}\left(
B_{qr}-N\delta _{qr}n_{q}\right) +N\delta _{ps}\delta _{qr}\left(
n_{p}-n_{q}\right)  \nonumber \\
n_{p} &=&<\psi _{p}^{+}\psi _{p}>=\mbox{Fermi function}  \nonumber
\end{eqnarray}
We change notation and use subtracted bilinears $B_{pq}\equiv \psi
_{p}^{+}\psi _{q}-N\delta _{pq}n_{p}$ from now on. The preceding equation
shows that the operators $B_{pq}$ for $p$, $q$ on {\bf opposite} sides of
the Fermi surface behave like harmonic oscillators: \ 
\begin{eqnarray}
B_{\overline{p},q} &\sim &\sqrt{N}b_{\overline{p},q}\mbox{ , \ \ \ }B_{p,%
\overline{q}}\sim \sqrt{N}b_{\overline{p},q}^{+}\mbox{ \ } \\
\mbox{\ with \ }[b_{\overline{p},q},b_{\overline{r},s}^{+}] &=&\delta _{%
\overline{p},\overline{r}}\delta _{q,s}  \nonumber
\end{eqnarray}
with, so far, unknown corrections. Above we denoted a level by $p$ when it
is empty and by $\overline{p}$ when it is full. Fortunately for us, the
expansion, to all orders, of bilinears of fermions in terms of bosons was
previously worked out in nuclear physics, see \cite{Klein+Marshalek} for a
review. Although we will not need this representation in its most general
form we quote it from ref \cite{Itakura}, according to which the leading
density operators are represented by 
\begin{eqnarray}
B_{\overline{p},q} &=&\sum_{r}\left( \sqrt{N-A}\right) _{q,r}b_{\overline{p}%
,r}  \label{bosonrep1} \\
\left[ b_{\overline{r},p},b_{\overline{s},q}^{+}\right] &=&\delta
_{p,q}\delta _{\overline{r},\overline{s}}  \nonumber \\
A_{pq} &=&b_{\overline{r},q}^{+}b_{\overline{r},p}  \nonumber
\end{eqnarray}
and where the square root $\sqrt{N-A}$ is to be interpreted as a series in $%
1/N$. By contrast, the non leading density operators, with momenta on the
same side of the Fermi surface, are simple quadratic forms in the harmonic
oscillators: 
\begin{eqnarray}
B_{pq} &=&\sum_{r}b_{\overline{r},p}^{+}b_{\overline{r},q}=A_{qp}
\label{finite} \\
B_{\overline{p},\overline{q}} &=&N-\sum_{r}b_{\overline{q},r}^{+}b_{%
\overline{p},r}  \nonumber
\end{eqnarray}
The representation of the algebra (\ref{heuristics}) via eqs(\ref{bosonrep1},%
\ref{finite}) constitutes a generalization of the Holstein-Primakoff
representation of eq(\ref{Holstein}), but the proof that it represents the
fermion bilinears correctly is much more difficult for this generalization
than\ for the Holstein-Primakoff representation\cite{KMProof}. Here we only
use the lowest order terms of this expansion and we retain as dynamical
variables the bilinears $B_{p\overline{q}}$ and $B_{\overline{p}q}$ with $p$%
, $q$ on {\bf opposite} sides of the Fermi surface where they have, from eq(%
\ref{heuristics}), harmonic oscillator like commutators

\begin{equation}
n_{p}+n_{q}=1:[B_{pq},B_{rs}]=N\delta _{ps}\delta _{qs}\left(
n_{p}-n_{q}\right) +O(1)  \label{streamlined1}
\end{equation}
The non leading $B_{pq}$ for $p$, $q$ on the {\bf same} side of the Fermi
surface are expressed, using eqs(\ref{finite}, \ref{bosonrep1}) in terms of
the leading operators as follows 
\begin{eqnarray}
n_{p} &=&n_{q}:  \label{streamlined2} \\
B_{pq} &=&\frac{1}{N}\sum_{r}\left( n_{r}-n_{p}\right) B_{pr}B_{rq}+O\left( 
\frac{1}{N}\right) +const  \nonumber
\end{eqnarray}
(our notation no longer distinguishes between filled and empty levels). We
may now use eq(\ref{streamlined2}) to express the kinetic energy in eq(\ref
{Fourier}) in terms of leading density operators

\begin{equation}
\sum_{p}e_{p}B_{pp}=\frac{1}{2N}\sum_{n(p)+n(q)=1}|e_{p}-e_{q}|\left\{
B_{pq}B_{qp}\right\} _{symm}+const+O\left( \frac{1}{N}\right)
\label{cinetic}
\end{equation}
where we have antisymmetrized $e_{p}$ with respect to the replacement $%
p\leftrightarrow q$. Using eq(\ref{cinetic}) the Hamiltonian in eq(\ref
{Fourier}) reads as follows: 
\begin{equation}
H_{SU(N)}=\frac{U}{2N\cdot vol}\sum_{p-q=r-s}B_{pq}^{+}B_{rs}+\frac{1}{2N}%
\sum_{n(p)+n(q)=1}|e_{p}-e_{q}|\left\{ B_{pq}B_{qp}\right\} _{symm}+const+O(%
\frac{1}{N})  \label{RPAlike}
\end{equation}
As we shall see later in more detail, this describes RPA like particle hole
scattering, provided we restrict the summation in the interaction term to $%
n(p)+n(q)=1$ and $n(r)+n(s)=1$ so that $B_{pq}^{+}B_{rs}\sim N$ .

The operators $B_{pq}^{+}B_{rs}$ in eq(\ref{RPAlike}) contain, however, also
non leading contributions with either $p,q$ or $r,s$ on the same side of the
Fermi surface, or both. By using eq(\ref{streamlined2}) to express these non
leading contributions in terms of the leading operators, we find, to next
order in $1/\sqrt{N}$, the following three point interaction: 
\begin{equation}
H^{(3)}=\frac{U}{N^{2}\cdot vol}\sum_{p-q=r-s;t}B_{qp}\left(
n_{t}-n_{r}\right) B_{rt}B_{ts}+const+O(N^{-3/2})
\end{equation}
The interaction $H^{(3)}$ is $\sim N^{-1/2}$ because it contains three
operators $\sim \sqrt{N}$ each with arguments on mutually opposite sides of
the Fermi surface and it gives rise to a correction $\sim \frac{1}{N}$ of
the energy. In this paper we will not make use of such non leading
interactions.

Finally we check whether the RPA like expression of eq(\ref{RPAlike})
reduces to our previous result for the dimer. We choose a naive bandwidth of
one so that $|e_{\pi }-e_{0}|=1$ and take into account $vol=2$ to obtain: 
\begin{equation}
H_{SU(N)}^{\mbox{dimer}}=\frac{U}{4}\left( b_{0\pi }+b_{0\pi }^{+}\right)
^{2}+b_{0\pi }^{+}b_{0\pi }+const+O(\frac{1}{N})
\end{equation}
It is reassuring that this agrees with our previous result (\ref{Holstein})
that was obtained directly via the Holstein Primakoff representation of
pseudo spin operators.

\section{Exact results for the symplectic Hubbard model}

\subsection{Map of the symplectic model onto a double exchange model}

The symplectic model of eq(\ref{symplectic}) is fairly complicated and to
make progress we first map it onto a spin model. A hint on the mapping we
need is provided by the fact that the operators \{$\psi _{x}\varepsilon \psi
_{x}$, $\left( \psi _{x}\varepsilon \psi _{x}\right) ^{+}$, $\psi
_{x}^{+}\psi _{x}$\} form an $SU(2)$ algebra under commutation. This fact
becomes obvious after a particle hole transformation $\psi _{x1}\rightarrow
\psi _{x1}$, $\psi _{x2}\rightarrow \psi _{x2}^{+}$ on the down spins: 
\begin{equation}
\left( 
\begin{array}{c}
\frac{1}{2}\psi _{x}^{+}\varepsilon \psi _{x}^{+} \\ 
-\frac{1}{2}\psi _{x}\varepsilon \psi _{x} \\ 
\frac{1}{2}\psi _{x}^{+}\psi _{x}
\end{array}
\right) \rightarrow \left( 
\begin{array}{c}
\psi _{x1}^{+}\psi _{x2} \\ 
\psi _{x2}^{+}\psi _{x1} \\ 
\frac{1}{2}\left( \psi _{x1}^{+}\psi _{x1}-\psi _{x2}^{+}\psi _{x2}\right)
\end{array}
\right) =\left( 
\begin{array}{c}
S_{x}^{+} \\ 
S_{x}^{-} \\ 
S_{x}^{3}
\end{array}
\right)  \label{SU2}
\end{equation}
We see that this particle hole transformation \cite{particlehole} replaces
the difficult charge $\pm 2$ pairing operators by $S_{x}^{\pm }$ while
density fluctuations get mapped onto $S_{x}^{3}$. Applying this map on the
symplectic Hamiltonian of eq(\ref{symplectic}) we find: 
\begin{equation}
H_{U(N,q)}\rightarrow H_{DE}=\frac{4U}{N}\sum_{x}\overrightarrow{S}_{x}\cdot 
\overrightarrow{S}_{x}+\sum_{x,y}t_{xy}\psi _{x}^{+}\sigma _{3}\psi _{y}
\label{de}
\end{equation}
While in the symplectic model of eq(\ref{symplectic}) up and down electrons
transform according to mutually conjugate representations under $SU(N)$ they
transform according to the same $SU(N)$ representation in the model of eq(%
\ref{de}) and therefore all four bilinears $\psi _{x\alpha }^{+}\psi
_{y\beta }$ become legitimate $SU(N)$ invariant operators. The Hamiltonian $%
H_{DE}$ is of the double exchange type, with only a single band and a Hund
coupling of the ''wrong sign'' that favors antiferromagnetic order and we
will refer to it as ''double exchange model'' for short. The chemical
potential $\psi _{x}^{+}\psi _{x}$ of the symplectic model transforms into $%
\psi _{x}^{+}\sigma _{3}\psi _{x}$ in the double exchange model where it
acts like a magnetic field that reduces the symmetry to spin rotations about
the $z$ axis \cite{ChoiceOfInteraction}.

\subsection{Asymptotic Hamiltonian of the double exchange model}

We first rewrite the double exchange model (\ref{de}) in terms of plane
waves as 
\begin{eqnarray}
H_{DE} &=&\frac{U}{Nvol}\sum_{p+r=q+s}\overrightarrow{\sigma }_{pq}\cdot 
\overrightarrow{\sigma }_{rs}+\sum_{p}e_{p}\rho _{pp}  \label{DEFourier} \\
\overrightarrow{\sigma }_{pq} &=&\psi _{p\alpha }^{+}\overrightarrow{\tau }%
_{\alpha \beta }\psi _{q\beta }\mbox{,
\ \ }\rho _{pq}=\psi _{p\alpha }^{+}\psi _{q\alpha }  \nonumber
\end{eqnarray}
where $\overrightarrow{\tau }_{\alpha \beta }$ are the Pauli matrices. The
fermion bilinears $B_{p\alpha ,q\beta }=\psi _{p\alpha }^{+}\psi _{q\beta }$
of the double exchange model now turn into RPA like collective modes in the
same way as the spinless $\psi _{p}^{+}\psi _{q}$ pairs did in the $SU(N)$
model of the last section, except for replacing $p\rightarrow (p,\alpha )$, $%
q\rightarrow (q,\beta )$ and $n_{p}\rightarrow n_{p\alpha }$ in eqs(\ref
{streamlined1},\ref{streamlined2}): 
\begin{eqnarray}
n_{p\alpha }+n_{q\beta } &=&1:[\psi _{p\alpha }^{+}\psi _{q\beta },\psi
_{r\gamma }^{+}\psi _{s\delta }]=N\delta _{pr}\delta _{qs}\delta _{\alpha
\delta }\delta _{\beta \gamma }\left( n_{p\alpha }-n_{q\beta }\right) +O(1)
\label{complicated} \\
n_{p\alpha } &=&n_{q\beta }:\mbox{ }B_{p\alpha ,q\beta }\mbox{\ }=\frac{1}{N}%
\sum_{r,\gamma }\left( n_{r\gamma }-n_{p\alpha }\right) B_{p\alpha ,r\gamma
}B_{r\gamma ,q\beta }+O\left( \frac{1}{N}\right) +const
\end{eqnarray}
The Fermi factors in the double exchange model depend on the spin direction
for non zero doping. This will break its $SU(2)\ast U(1)$ symmetry and split
the multiplet of modes. To simplify our analysis, we therefore limit
ourselves here to zero doping for which the Fermi surfaces in the double
exchange model loose their spin dependence. We may expand the collective
modes $\psi _{p\alpha }^{+}\psi _{q\beta }$ of eq(\ref{complicated}) over
the Pauli matrices $\overrightarrow{\tau }$ and decompose\ them into scalar $%
\rho _{pq}$ and vector \ $\overrightarrow{\sigma }_{pq}$ operators according
to 
\begin{eqnarray}
\psi _{p\alpha }^{+}\psi _{q\beta } &=&\frac{1}{2}\sum_{\mu =0..3}\sigma
_{pq}^{\mu }\tau _{\beta \alpha }^{\mu }\mbox{, \ }\tau ^{\rho }=\left( 1,%
\overrightarrow{\tau }\right)  \label{decomposition} \\
\sigma _{pq}^{\mu } &=&\left( \rho _{pq},\overrightarrow{\sigma }%
_{pq}\right) \mbox{\ }  \nonumber
\end{eqnarray}
Using eqs(\ref{complicated},\ref{decomposition}) the modes $\rho _{pq},%
\overrightarrow{\sigma }_{pq}$ can be shown to be independent degrees of
freedom, to leading order in $N$: 
\begin{equation}
\left[ \sigma _{p,q}^{\mu },\sigma _{r,s}^{\mu }\right] =2N\delta _{\mu \nu
}\delta _{ps}\delta _{qr}\left( n_{p}-n_{q}\right) +O(1)  \label{O4}
\end{equation}
Use of this representation on eq(\ref{DEFourier}) for $H_{DE}$ gives us the
RPA dynamics of $\rho _{p,q}$ and $\overrightarrow{\sigma }_{p,q}$ in the
undoped DE model: 
\begin{eqnarray}
H_{DE} &=&\frac{U}{N\ast vol}\sum_{p+s=q+t}\overrightarrow{\sigma }%
_{pq}\cdot \overrightarrow{\sigma }_{st}+\frac{1}{4N}\sum_{p,q}\left(
e_{p}-e_{q}\right) \left( n_{q}-n_{p}\right) \left[ \rho _{pq}\rho _{qp}+%
\overrightarrow{\sigma }_{pq}\cdot \overrightarrow{\sigma }_{qp}\right]
\label{rpaDE} \\
&&+const+O\left( \frac{1}{N}\right)  \nonumber
\end{eqnarray}
where we antisymmetrized $e_{p}$ with respect to interchange $%
p\leftrightarrow q$. $H_{DE}$ contains also nonlinear interactions in $%
\overrightarrow{\sigma }_{pq}\cdot \overrightarrow{\sigma }_{st}$ with
either $p$, $q$ or $r$, $s$ on the same side of the Fermi surface, but we
will not write them down here. From our previous discussion we know that the
operators $S^{\pm }$ , $S^{3}$ correspond, respectively, to chargedand
neutral modes and, similarly, $\overrightarrow{\sigma }_{pq}$ decomposes
into charged ($\sigma _{pq}^{\pm }$) and neutral ($\sigma _{pq}^{3}$)
bosons. The basic reason for the appearance of charged collective modes in
the symplectic model of eq(\ref{symplectic}) is the interaction of pairs in
its Hamiltonian.

\subsection{Solution, for arbitrary $N$, of the symplectic model on a dimer
of two points.}

We consider the dimer both in its symplectic and its double exchange version
by specializing eqs(\ref{symplectic},\ref{de}) to two points: 
\begin{eqnarray}
H_{SY}^{\mbox{dimer}} &=&\frac{U}{N}\sum_{x=1,2}\left[ \left\{ \left( \psi
_{x}\varepsilon \psi _{x}\right) ^{+}\left( \psi _{x}\varepsilon \psi
_{x}\right) \right\} _{symm}+\left( \psi _{x}^{+}\psi _{x}\right) ^{2}\right]
+\frac{1}{2}\left( \psi _{1}^{+}\psi _{2}+\psi _{2}^{+}\psi _{1}\right)
\label{dimer} \\
H_{DE}^{\mbox{dimer}} &=&\frac{4U}{N}\left( \overrightarrow{S}_{1}\cdot 
\overrightarrow{S}_{1}+\overrightarrow{S}_{2}\cdot \overrightarrow{S}%
_{2}\right) +\frac{1}{2}\left( \psi _{1}^{+}\psi _{2}+\psi _{2}^{+}\psi
_{1}\right)  \nonumber
\end{eqnarray}
For a dimer of two points, the Hamiltonian of eq(\ref{rpaDE}) reduces to 
\begin{equation}
H_{DE}^{\mbox{dimer}}=\frac{U}{2N}\left( \overrightarrow{\sigma }_{0\pi }+%
\overrightarrow{\sigma }_{\pi 0}\right) ^{2}+\frac{1}{2N}|e_{\pi }-e_{0}|%
\left[ \rho _{0\pi }\rho _{\pi 0}+\overrightarrow{\sigma }_{0\pi }%
\overrightarrow{\sigma }_{\pi 0}\right] +const+O(\frac{1}{N})
\label{SymplecticDimerRPA}
\end{equation}
According to eq(\ref{O4}) the operators $\sigma _{0\pi }^{\mu }=\left\{ \rho
_{0\pi },\overrightarrow{\sigma }_{0\pi }\right\} $ may be represented as $%
\sigma _{0\pi }^{\mu }=\sqrt{2N}b^{\mu }$ to leading order and therefore the
symplectic dimer with naive bandwidth $|e_{\pi }-e_{0}|=1$ \ at $N=\infty $
is represented in terms of harmonic oscillators as follows: 
\begin{equation}
\lim_{N\rightarrow \infty }H_{DE}^{\mbox{dimer}}=\frac{U}{2}\left( 
\overrightarrow{b}+\overrightarrow{b}^{+}\right) ^{2}+\left( \overrightarrow{%
b}^{+}\overrightarrow{b}+b_{0}^{+}b_{0}\right)  \nonumber
\end{equation}
By referring to eqs(\ref{Holstein}, \ref{Bogoljubov}) for the $SU(N)$ dimer
we can read off the frequencies of the oscillators after Bogoljubov
transformation and we find $\omega _{0}=1$, $\omega _{1..3}=\sqrt{1+2U}$.

Straightforward numerical diagonalization of the dimer is impossible because
its number of states grows as $16^{N}$ with $N$. However, according to exact
diagonalization for $N=1,2,3$, the low lying states of the dimer are
included in a much smaller subspace of $SU(N)$ invariant states. To further
simplify the diagonalization within the $SU(N)$ invariant subspace, it is
useful to recognize that the building blocks of the symplectic dimer $\psi
_{a}\varepsilon \psi _{b}$, $\psi _{a}^{+}\psi _{b}$ for\ $a,b=1,2$ in eq(%
\ref{dimer}) generate the algebra of $SP(4)\sim SO(5)$. This can be seen by
choosing two commuting operators from this set such as $\psi _{1}^{+}\psi
_{1}$ and $\psi _{2}^{+}\psi _{2}$ and by determining the Cartan weights
under commutation with the remaining operators and by plotting the two
dimensional weight diagram. Opening a textbook on group theory and comparing
with the Cartan weight diagrams of the classical groups of low rank \cite
{CartanWeight} one concludes that the algebra is $SP(4)\sim SO(5)$.

The irreducible representations of $SO(5)$ have been classified with respect
to their $SU(2)\times SU(2)$ content and we only need the special
irreducible representation that contains the vacuum state of the double
exchange model and which is characterized by $S_{1}=S_{2}=\frac{k}{2}$, $%
k=0...N$. The kinetic term in this representation can then be determined
from the reduced matrix elements given in \cite{Kemmer}.

\begin{center}
\begin{figure}[h]
\mbox{\epsfig{file=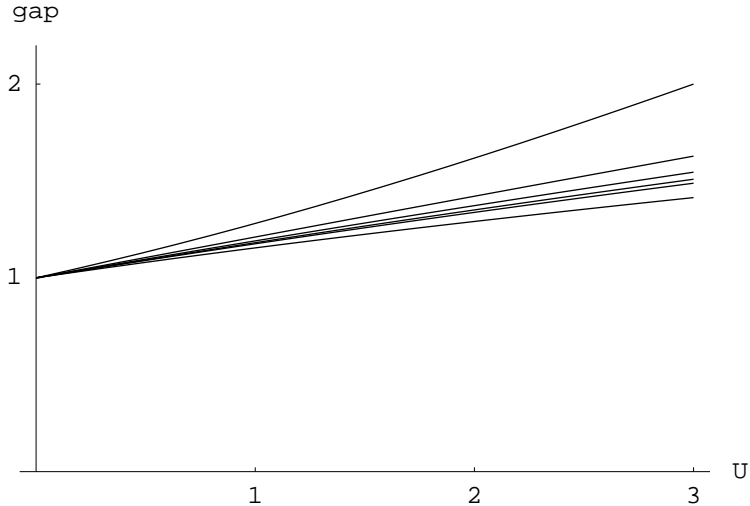,width=10cm}}\\
\caption{Monotonic descent and convergence of the gaps $\Delta _{N}(U)$ of
the symplectic dimer for N=1..5 towards $\Delta _{\infty }(U)$. The highest
curve is for $N=1$, the lowest curve represents $N=\infty $.}
\end{figure}
\end{center}

From numerical diagonalization of the resulting representation of the
Hamiltonian that we consider at zero doping we find the first excited state
of the dimer to be a degenerate vector triplet. In order for $U$ to have its
conventional meaning at $N=1$ we replaced $U\rightarrow \frac{U}{6}$ in all
formulas and plotted the energy $\Delta _{N}(U)$ of the vector multiplet
relative to the ground state in figure {\bf 2} for $N=1..5$ together with
its asymptotic limit $\Delta _{\infty }(U)=\sqrt{1+U/3}$. \ Comparing
figures {\bf 1} and {\bf 2} we conclude that convergence of the gaps towards
their asymptotic limit is slower for the symplectic extrapolation than it
was for $SU(N)$.

\section{ The spectrum of the $SU(\infty )$ and $U(\infty,q) $ Hamiltonians}

Although eqs(\ref{RPAlike},\ref{rpaDE}) give the Hamiltonians of the $%
SU(\infty )$ and $U(\infty ,q)$ models in the sector of even Fermion number
in terms of elementary harmonic oscillators the spectra of these
Hamiltonians are not obvious at first sight. This is due to a serious
overcounting of degrees of freedom, with roughly one boson for any pair of
momenta or any pair of points. These essentially non local bosons can
describe local physics only by virtue of the interaction $\sim U$ being
local.

We wish to determine the spectra of the asymptotic Hamiltonians via the
poles of some correlation functions and begin with the correlator $%
<B_{p_{1}q_{1}}B_{p_{2}q_{2}}^{+}>$ of the $SU(\infty )$ theory for
simplicity. As a further simplification, we adopt the more flexible
functional integral representation of the partition function which is fairly
obvious here because we are dealing with a collection of independent
oscillators: 
\begin{eqnarray}
B_{pq} &=&\sqrt{N}b_{pq} \\
n_{p}+n_{q} &=&1:[b_{pq},b_{rs}^{+}]=\delta _{pr}\delta _{qs}\left(
n_{p}-n_{q}\right) +O(1/N)  \nonumber
\end{eqnarray}
(we have rescaled the oscillators to have conventional commutation
relations). The Lagrangian we need is just a slight generalization of the
conventional one: 
\begin{eqnarray}
L &=&\frac{1}{2}\sum_{p,q}\left( n_{p}-n_{q}\right) b_{pq}^{+}\partial
_{t}b_{pq}-h_{SU(N)}(b,b^{+}) \\
Z &=&\int DBe^{-S}\mbox{ with \ }S=\int_{0}^{\beta }dtL=b^{+}Mb  \nonumber
\end{eqnarray}
Correlators (in imaginary time) of harmonic variables such as $b$ and $b^{+}$
are well known to be given by the inverse of the quadratic form in the
imaginary time action: 
\begin{equation}
<b_{p_{1}q_{1}}(t)b_{p_{2}q_{2}}^{+}(0)>=<p_{1}q_{1};t|\frac{1}{{\it M}}%
|p_{2}q_{2};0>
\end{equation}
At this point we invoke the locality of the interaction $\sim U$ \ which, in 
$x$ space and in terms of the rescaled variables $b$, reads: 
\begin{eqnarray}
\int_{0}^{\beta }dth_{int} &=&U\int_{0}^{\beta
}dt\sum_{x}b_{xx}^{+}b_{xx}=U\int_{0}^{\beta
}dt\sum_{x}b_{xy}^{+}P_{xy,x^{\prime }y^{\prime }}b_{x^{\prime }y^{\prime }}
\\
\mbox{with }P_{xy,x^{\prime }y^{\prime }}b_{x^{\prime }y^{\prime }}
&=&\delta _{xy}b_{xx}  \nonumber
\end{eqnarray}
The operator $P$ picks out the coincident piece of the wave function and
satisfies the projector relation $P^{2}=P$. We decompose the full action
into its ''free''and ''interacting'' part by writing $%
b^{+}Mb=b^{+}Lb-Ub^{+}Pb$ with 
\begin{equation}
b^{+}{\it L}b=\int dt\left[ \frac{1}{2}\sum_{p,q}\left( n_{p}-n_{q}\right)
b_{pq}^{+}\partial _{t}b_{pq}-\frac{1}{2}\sum_{n(p)+n(q)=1}\left(
n_{p}-n_{q}\right) (e_{p}-e_{q})\left\{ b_{pq}^{+}b_{pq}\right\} _{symm}%
\right] 
\end{equation}
where ${\it L}$ describes free propagation of particle hole pairs with $U=0$%
. According to general principles, the free propagator $<b(t)b^{+}(0)>$ is
just the inverse of ${\it L}$ : 
\begin{equation}
<b_{pq}(t)b_{rs}^{+}(0)>_{U=0}=\delta _{pr}\delta _{qs}<pq;t|\frac{1}{{\it L}%
}|rs;0>=\delta _{pr}\delta _{qs}\frac{n_{p}-n_{q}}{\partial _{t}-\left(
e_{p}-e_{q}\right) }
\end{equation}
The last expression resembles the particle hole propagator that enters into
the calculation of bubbles and Lindhard's function and this indicates that
we are on the right track. Next we switch on $U$ and exploit $P^{2}=P$, the
locality of the interaction, in a way reminiscent of the treatment of a
Friedel resonance: 
\begin{eqnarray}
&<&b(t)b^{+}(0)>_{U\neq 0}=\frac{1}{{\it M}}=\frac{1}{{\it L}-U{\it P}} \\
&=&\frac{1}{{\it L}}+U\frac{1}{{\it L}}{\it P}\frac{1}{1-U{\it P}\frac{1}{%
{\it L}}{\it P}}{\it P}\frac{1}{{\it L}}  \nonumber
\end{eqnarray}
${\it P}\frac{1}{1-U{\it P}\frac{1}{{\it L}}{\it P}}{\it P}$ represents
propagation and rescattering \ of particle hole pairs or a sum of bubbles.
The preceding argument shows that the spectrum of the $SU(\infty )$
Hamiltonian corresponds to zeroes of the operator 
\begin{equation}
\left( 1-U{\it P}\frac{1}{{\it L}}{\it P}\right) _{tx,0y}=\delta _{xy}\delta
(t)-U<xt|\frac{1}{{\it L}}|y0>
\end{equation}
The locality of the interaction $\sim U$ leads to a propagator of coincident
pairs and to functions of a single label. Fourier transforming we find 
\begin{eqnarray}
&<&xt|\frac{1}{{\it L}}|\overrightarrow{0}0>=<b_{xx}(t)b_{00}^{+}(0)> \\
&=&\frac{1}{V^{2}}\sum_{p,q}e^{i(p-q)x}\frac{n_{p}-n_{q}}{\partial
_{t}-\left( e_{p}-e_{q}\right) }=\frac{1}{\beta V^{2}}\sum_{p,q,\omega
}e^{i(p-q)x}e^{i\omega t}\frac{n_{p}-n_{q}}{i\omega -\left(
e_{p}-e_{q}\right) }  \nonumber
\end{eqnarray}
But the correlation $<b_{xx}(t)b_{00}^{+}(0)>$ for free fields is also
calculable via Wicks theorem: 
\begin{eqnarray}
&<&b_{xx}(t)b_{00}^{+}(0)>+O(\frac{1}{N})=<\left( \psi _{x}^{+}\psi
_{x}\right) _{t}\left( \psi _{0}^{+}\psi _{0}\right) _{0}>_{%
\mbox{no
internal label}}\equiv \chi (x)=-G(x)G(-x) \\
\mbox{with \ \ }G(x) &=&-<\psi _{x}(t)\psi _{0}^{+}(0)>=-\frac{1}{\partial
_{t}+h^{hop}}=\frac{1}{\beta V}\sum_{\omega ,p}\frac{e^{i(\omega t+px)}}{%
i\omega -e(p)}  \nonumber
\end{eqnarray}
where $h^{hop}$ is the hopping matrix. Using conventional Matsubara
techniques it is easy to check that 
\begin{equation}
\chi (x)=-G(x)G(-x)=\frac{1}{\beta V^{2}}\sum_{Q}e^{-iQx}\sum_{p-q=Q}\frac{%
n_{p}-n_{q}}{iq_{0}-(e_{p}-e_{q})}
\end{equation}
We conclude that the spectrum of the boson Hamiltonian corresponds to the
singularities of 
\begin{equation}
\frac{1}{\delta _{xy}\delta (t)-U<xt|\frac{1}{L}|y0>}=\frac{1}{\delta
_{xy}\delta (t)-U\chi (\overrightarrow{x},t)}
\end{equation}
where $\chi (x)=<\left( \psi _{x}^{+}\psi _{x}\right) \left( \psi
_{0}^{+}\psi _{0}\right) >$ is a susceptibility. At half filling, in d=2
dimensions and at $Q=(\pi ,\pi )$ the Fourier transform of $\chi (%
\overrightarrow{x},t)$ is well known \cite{Hirsch} to diverge
logarithmically to $+\infty $ \ because of an incipient antiferromagnetic
instability. We conclude that the asymptotic Hamiltonian $H_{SU(\infty )}$
describes bubbles that have a pole near $Q=(\pi ,\pi )$ at half filling.
This pole in the spin fluctuation channel is also seen in neutron scattering
on cuprates \cite{Aeppli}.

At half filling the spectrum of the asymptotic Hamiltonian $H_{U(\infty ,q)}$
eq(\ref{rpaDE}) on an infinite lattice\ is very similar to that of the $%
SU(\infty )$ model. Again the Hamiltonian describes bubbles and again the
spectrum is dominated by fluctuations at $Q=(\pi ,\pi )$ except that there
is now an extra band index. The excitations form a triplet of spin and pair
fluctuations that are exactly degenerate at half filling.

\section{Extrapolating from $N=\infty $ to $N=1$}

The aim of the present paper was to establish exact \ results on the
symplectic model in order to see whether the expansion in powers of $1/N$
about $N=\infty $ in this model makes any sense. The boson expansion helped
us find the asymptotic Hamiltonian for a finite number of points but as we
have seen in the last section this method becomes inconvenient for dealing
with infinite lattices where the bosons seriously overcount the degrees of
freedom. In this section we return to main stream methods of condensed
matter physics and discuss how to extrapolate to from $N=\infty $ to $N=1$.

From the preceding sections it follows that a treatment of doped
antiferromagnets and the Hubbard model at low temperature near half filling
requires a self consistent treatment of electron propagation and spin and
pair fluctuations. This may be done either using the topological expansion
given in \cite{DF} or by using the more conventional saddle point method.
For simplicity we adopt the saddle point method, although the topological
expansion is presumably the more powerful approach. We may set up a $1/N$
expansion of the double exchange model by conventional Hubbard-Stratonovich
transformation:

\begin{eqnarray}
\int D\phi \exp -\frac{U}{N}\left( \overrightarrow{\phi }_{x}-i%
\overrightarrow{s}_{x}\right) ^{2} &=&const\mbox{, \ \ \ }\overrightarrow{s}%
_{x}=\frac{1}{2}\psi _{\alpha i}^{\ast }\sigma _{\alpha \beta }\psi _{\beta
i} \\
Z &=&\int D\phi D\psi \exp -S\mbox{, \ \ \ }S=\int dtL  \nonumber \\
L &=&\sum_{x}N\overrightarrow{\phi }_{x}^{2}-2i\sqrt{U}\cdot \overrightarrow{%
s}_{x}\overrightarrow{\phi }+\psi ^{\ast }\left( \partial
_{t}+t^{hop}\right) \psi +\mu \psi ^{\ast }\sigma _{3}\psi  \nonumber
\end{eqnarray}
where $t^{hop}$ is the hopping amplitude. Like in preceding work on the $%
SU(N)$ Hubbard model at large $N$ (\cite{Affleck}) the phase factor ''$i$''
is essential to get the correct coefficient \ $+U$ for the coupling in the
double exchange model and again the factor ''$i$'' causes no difficulty
because $\overrightarrow{\phi }$ is purely imaginary at the saddle point. To
see this, we integrate over the fermionic degrees of freedom and find the
stationary point of the effective action:: 
\begin{eqnarray}
Z &=&\int D\overrightarrow{\phi }D\psi \exp -S=\int D\phi \exp -S_{eff} \\
\frac{S_{eff}(\overrightarrow{\phi })}{N} &=&\int \sum_{x}\overrightarrow{%
\phi }_{x}^{2}dt-\log \det \left( \partial _{t}+t^{hop}+\mu \sigma _{3}-i%
\sqrt{U}\cdot \overrightarrow{\sigma }\overrightarrow{\phi }\right) +const 
\nonumber \\
\delta S\bigskip _{eff} &=&0\rightarrow i\overrightarrow{\phi }(x,t)=\frac{1%
}{2}\sqrt{U}Tr\overrightarrow{\sigma }<xt|\frac{1}{\partial _{t}+t^{hop}+\mu
\sigma _{3}-i\sqrt{U}\cdot \overrightarrow{\sigma }\overrightarrow{\phi }}%
|xt>=\frac{\sqrt{U}}{N}<\overrightarrow{S}>  \nonumber
\end{eqnarray}
(there are $N$ species that contribute to $<\overrightarrow{S}>$, hence the
factor $1/N$). So the hypothesis that $i\overrightarrow{\phi }$ is real is
consistent, but difficulties may be anticipated for $\mu =i\sqrt{U}\phi ^{3}$%
. Numerical calculations indicate that the spin fluctuations are quickly
killed by doping while the pair fluctuation channel remains intact. The
likely explanation is that the spin fluctuations correspond to longitudinal
fluctuations of $\overrightarrow{\phi }$ in the $3$ direction while the pair
fluctuations correspond Goldstone like rotations about the $z$ axis.

One may also expand $S_{eff}$ to second order in $\overrightarrow{\phi }$
and confirm the existence of degenerate spin and pair fluctuations at half
filling and $N=\infty $ without ever using boson operators (but we do not
see how to avoid them in the dimer problem). A detailed study of the
extrapolation to $N=1$ via coupled integral equations will be given in a
separate \ paper.

\section{Conclusions}

In this paper we have derived two exact results that serve as first steps
towards a $1/N$ expansion of the symplectic Hubbard model. Firstly, we
diagonalized the symplectic dimer for any $N$ using an underlying $SP(4)\sim
SO(5)$ classifying symmetry. We found a smooth interpolation, without any
singularities, between $N=1$ and $N=\infty $ which suggests that the
expansion, in powers of $1/N$, about the point $N=\infty $ is indeed
legitimate.

Secondly, we determined the Hamiltonian of the symplectic model for $%
N=\infty $ and found a multiplet of collective modes containing charged pair
fluctuations and neutral spin fluctuations that are exactly degenerate with
each other at half filling. This tells us that pair fluctuations in doped
antiferromagnets are essential degrees of freedom

On the way to establishing these results we found analogous results for the $%
SU(N)$ extrapolation of the Hubbard model and we also indicated how
Holstein-Primakoff like boson operators may be replaced by more familiar and
more powerful Green's function techniques. The map of the original
symplectic model onto a magnetic model that is reminiscent of a double
exchange model proved crucial to our arguments. This map is interesting in
its own right because it may provide us with a simple physical picture of
the superconducting instability.

The extrapolation from $N=\infty $ to $N=1$ requires a self consistent
treatment of fermions interacting with their own triplet of collective modes
and taking into account mutual renormalization of bosonic and fermionic
fluctuations in a way similar to the FLEX approach. In the light of our
results, the FLEX approach is more closely related to the $SU(N)$
extrapolation of the Hubbard model than the $U(N,q)$ one as it ignores the
pairing fluctuations of the latter model. The integral equations of the
extrapolation to $N=1$ follow essentially from the topological
classification given previously in \cite{DF}, but the details remain to be
worked out. 

Several scenarios for the physics of pair fluctuations have been
suggested, ranging from preformed pairs \cite{PreformedPairs} to postulating
a quantum phase transition at a critical doping \cite{quantumphase}. Also,
various authors have added pair fluctuations on top of antiferromagnetic
ones and obtained interesting results \cite{Yamada}. It is fair to say,
however, that the details of the competition between spin and pairing
fluctuations still remain to be understood.

\noindent {\bf Acknowledgements. }We are crucially indebted to Peter Fulde
for his continued interest in this work and for the impetus he gave to this
project by inviting us to Max-Planck-Institut f\"{u}r Physik komplexer
Systeme at Dresden during a leave of D. F. from the University of Bordeaux.
D. F. is indebted to P. Stamp, G. 't Hooft, C. Lhuillier and P. Schuck for
providing support for this leave and making it possible in the first place.
We acknowledge helpful comments and/or encouragement by many colleagues, in
particular H. Beck, A. Georges, N. Hambli, M. Katsnelson, J. Ranninger, D.
Rowe and J.-M. Robin. We are indebted to A. Lichtenstein for criticism of an
earlier version of this paper and advice and code, to S. C. Wang for help
with preliminary numerics and to R. Hayn for discussions on the manuscript.


\begin{references}
\bibitem{U1U2}  The coefficients of the two interaction terms in the
symplectic model could be different in principle, but in \cite{DF} it is
shown why they should be equal.
\bibitem{SUNinvariance}  To have $SU(N)$ invariance we need $s=N/2$. With a
Hamiltonian corresponding more closely to tunnelling between quantum dots,
the optimal value of $s$ depends on $U$ and the charge is squeezed out of
the dimer for $U>1$.
\bibitem{particlehole}  This transformation is also used in Nambu's two
component formalism for anomalous Greens functions.
\bibitem{ChoiceOfInteraction}  The $O(3)$ symmetry in the double exchange
model without doping is due to the equality of the two coefficients in the
interaction in eq(\ref{symplectic}).
\end{references}
\end{document}